\documentclass{PoS}

\title{Fluctuations and correlations  from NA61/SHINE}

\ShortTitle{On fluctuations from NA61/SHINE}

\author{\speaker{Marek Gazdzicki}
	\\
	Goethe-Univeristy Frankfurt am Main,\\ 
	Jan Kochanowski University Kielce\\
	E-mail: \email{marek@cern.ch}}

\author{for the NA61/SHINE Collaboration }
\abstract{ 
          Basic ideas, methods and results related to the NA61/SHINE study of event-by-event fluctuations 
          in high energy nuclear collisions are briefly reviewed.
  		}

\FullConference{Critical Point and Onset of Deconfinement - CPOD2017\\
		7-11 August, 2017\\
		The Wang Center, Stony Brook University, Stony Brook, NY}

\begin{document}

\section{Introduction}

Study of event-by-event fluctuations is the focus of the NA61/SHINE programme on strong interactions.
Initially the study was motivated by the possibility to discover the critical point of strongly interacting matter
and a need to understand how the onset of deconfinement influences event-by-event 
fluctuations~\cite{na61proposal}.
Recently, results on fluctuations together with data on mean hadron multiplicities allowed to
uncover the onset of fireball~\footnote{The name was proposed by Edward 
Shuryak during the CPOD~2017 workshop in Stony Brook} -  
the rapid change of hadron production properties that start when moving from Be+Be 
to Ar+Sc collisions~\cite{sr2017}. 

Fluctuations in high energy collisions are significantly influenced by fluctuations in the amount of matter
(\textit{volume}) and energy involved in a collision, 
as well as global and local conservation laws.
In the search for the critical point and the study of the onset of deconfinement these are
unwanted effects~\cite{critical}. However, the onset of fireball was discovered thanks to the sensitivity of 
fluctuations to the conservation laws~\cite{sr2017}. 
Moreover, the dynamics of the early stage of the collision can be studied exploiting
\textit{volume} fluctuations~\cite{mixing}.

In this contribution \textit{volume} fluctuations are treated as unwanted effect and methods to remove their
influcence are presented in Secs.~\ref{sec:volume} and~\ref{sec:critical}.
The conservation laws are in Sec.~\ref{sec:conservation} treated as a tool to uncover the onset of fireball
and in Sec.~\ref{sec:critical} as to-be-minimized bias in the search for the critical point.

\section{Removing \textit{volume} fluctuations}
\label{sec:volume}

\begin{figure}[h]
	\centering
	\includegraphics[width=0.50\textwidth]{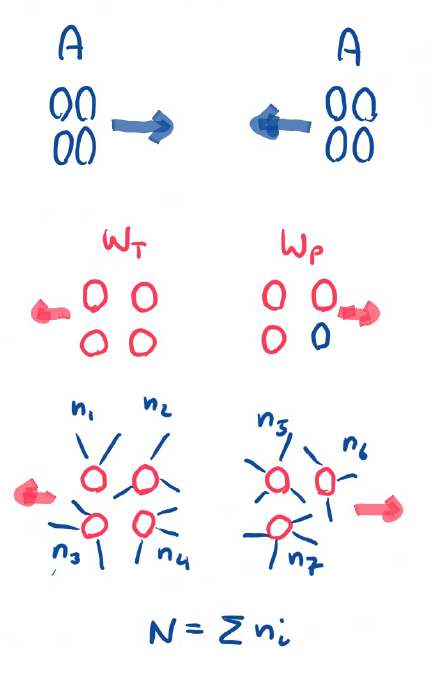}
	\caption[]
	{
		Sketch of particle production in nucleus-nucleus collisions according to the Wounded Nucleon Model.
		Projectile and target nuclei with nuclear mass number $A_P$ and $A_T$ (here $A = A_P =A_T = 4$)
		collide. $W_T$ (here $ W_T = 4 $) target wounded nucleons and $W_P$ (here $ W_P = 3 $)
		projectile wounded nucleons produce $N$ particles, where $N$ is given by the sum over all 
                wounded nucleons of particle multiplicities $n$ from single wounded nucleons.
	}
	\label{fig:wnm}
\end{figure}

It is probably the simplest to introduce fluctuations of the amount of matter involved in a collision
and their impact on fluctuations of produced particles using as an example 
the Wounded Nucleon Model~\cite{wnm}.
The model was proposed in 1976
as a late child of the S-matrix period~\cite{history}.
It assumes that particle production in nucleon-nucleon 
and nucleus-nucleus collisions is an incoherent superposition 
of particle production from wounded nucleons (nucleons 
which interacted inelastically and whose number is calculated using straight line trajectories of nucleons).
Properties of wounded nucleons are independent of the size of colliding
nuclei, e.g. they are the same in p+p and Pb+Pb collisions at the same
collision energy per nucleon.
These assumptions are graphically illustrated in Fig.~\ref{fig:wnm}.

Let us consider multiplicity (particle number) fluctuations characterized
by second moments of multiplicity distributions.
Two quantities of relevance are variance,
$Var[N] = \langle (N - \langle N \rangle)^2 \rangle$ and
scaled variance, $\omega[N] = Var[N]/\langle N \rangle$.
Here $N$ and $\langle N \rangle$ stand for multiplicity and its mean value, respectively.
Then for any probability distribution $P(W)$,
the scaled variance calculated within WNM reads~\cite{siq}:
\begin{equation}
\omega[N] = \omega[N]_W + \langle N \rangle / \langle W \rangle \cdot \omega[W] ~, \\ 
\label{eq:wnm:varN}
\end{equation}
where $\omega[N]_W$ stands for the scaled variance at any fixed number of wounded nucleons, and
$W = W_P + W_T$ for the sum of the number $W_P$ and $W_T$ of projectile and target nucleons.
Here the first component of Eq.~\ref{eq:wnm:varN} is considered to be the wanted one, whereas
the second one is unwanted.
Similar relations are valid for Statistical Models of an Ideal Boltzmann gas within the Grand Canonical Ensemble
SM(IB-GCE)~\cite{siq}.   

\begin{figure}[h]
	\centering
	\includegraphics[width=0.90\textwidth]{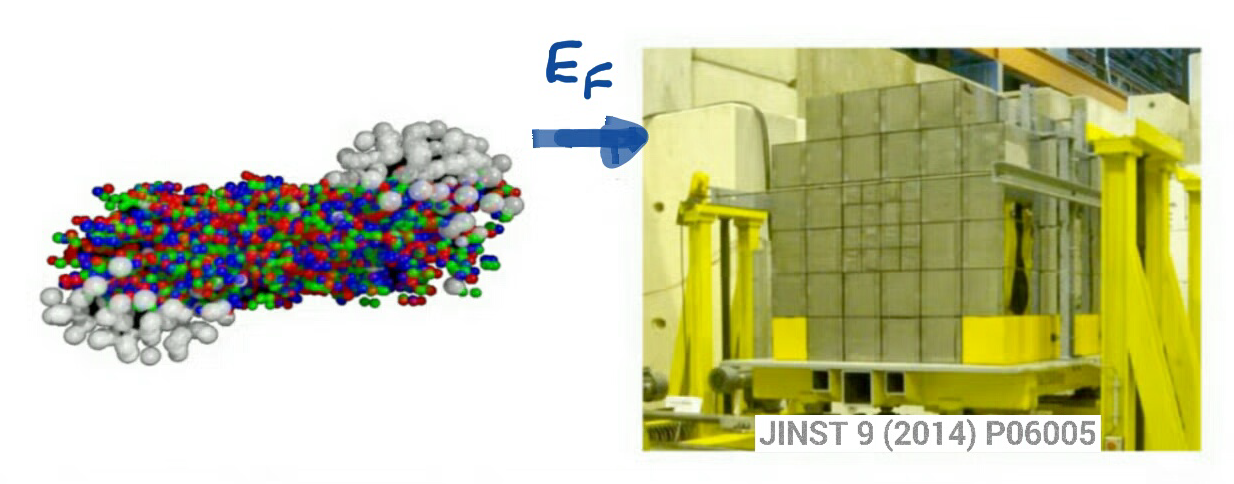}
	\caption[]
	{
		Energy $E_F$ recorded in a set of modules of the Projectile Spectator Detector is used to select 
		the most violent nucleus-nucleus collisions.
		$E_F$ is dominated by the energy of projectile spectators. 
	}
	\label{fig:psd}
\end{figure}

\begin{figure}[h]
	\centering
	\includegraphics[width=0.90\textwidth]{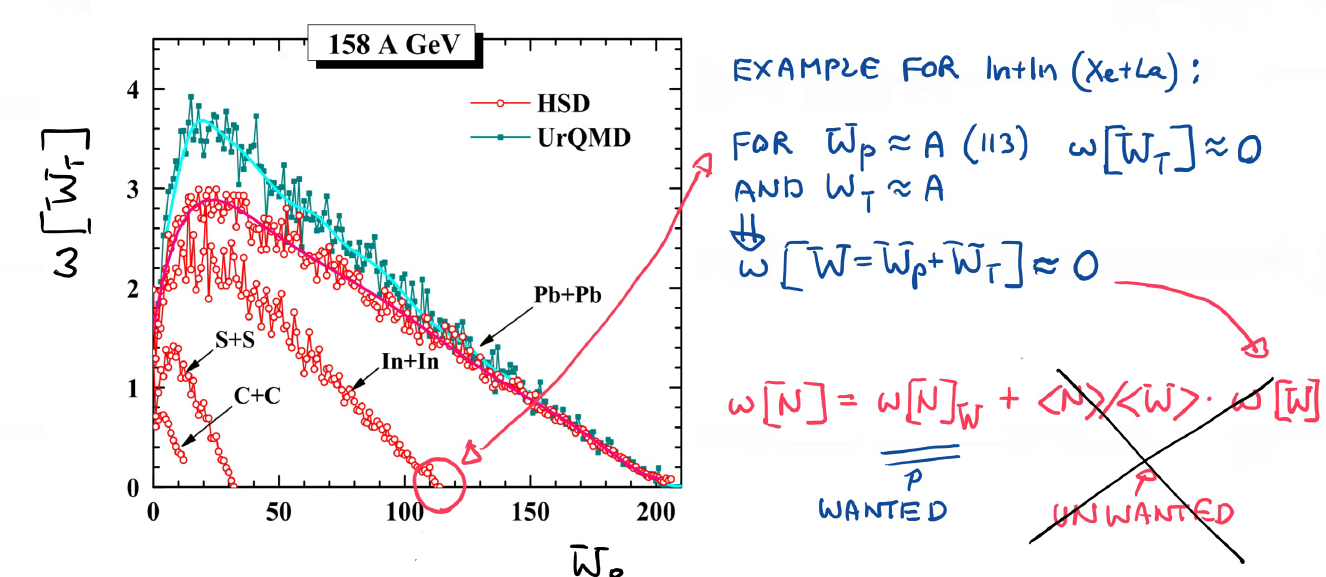}
	\caption[]
	{
		Scaled variance of the dsitribution of the number $W_T$ of wounded target nucleons as a 
                function of the number $W_P$ of wounded projectile 
		nucleons calculated with the HSD and UrQMD models for different nucleus-nucleus collisions. 
	}
	\label{fig:omega_W}
\end{figure}

In order to limit the unwanted component NA61/SHINE selects collisions with the smallest 
energy recorded by the Projectile Spectator Detector - the calorimeter that predominantly measures
energy of projectile spectators, see Fig.~\ref{fig:psd}. 
Multiplicity fluctuations for the selected most violent collisions 
(collisions with the largest $W_P$) are only weakly increased by $W$ fluctuations. This is demonstrated
in Fig.~\ref{fig:omega_W} based on simulations performed with the HSD and UrQMD models~\cite{hsdurqmd}.

Since even for the most violent collision volume fluctuations 
cannot be fully eliminated, it is important
to further minimise their effect by defining suitable fluctuation measures.
It appears that using second and first moments of the distribution of two extensive
quantities 
(their first moments are proportional to  \textit{volume})
one can construct fluctuation measures which are, 
for the WNM and the SM(IB-GCE) models~\cite{mgsm,siq,sangaline},
independent of \textit{volume} fluctuations.

In particular, one can construct the strongly intensive scaled variance~\cite{siq},
\begin{equation}
\Omega[N,E_P] = \omega[N] - ( \langle N \cdot E_P \rangle - \langle N \rangle \cdot \langle E_P \rangle ) / 
\langle E_P \rangle~,  
\label{eq:Omega}
\end{equation}
where $E_P = E_{BEAM} - E_F$.
It is easy to show that
\begin{equation}
\Omega[N,E_P] \approx \omega[N]_W = \omega[n]
\end{equation}
under two conditions $\langle N \cdot E_P \rangle_W \approx \langle N \rangle_W \cdot \langle E_P \rangle_W$ and
$\langle E_P \rangle_W \sim W$ that are expected to be fulfilled for violent A+A collisions.

\begin{figure}[h]
	\centering
	\includegraphics[width=0.90\textwidth]{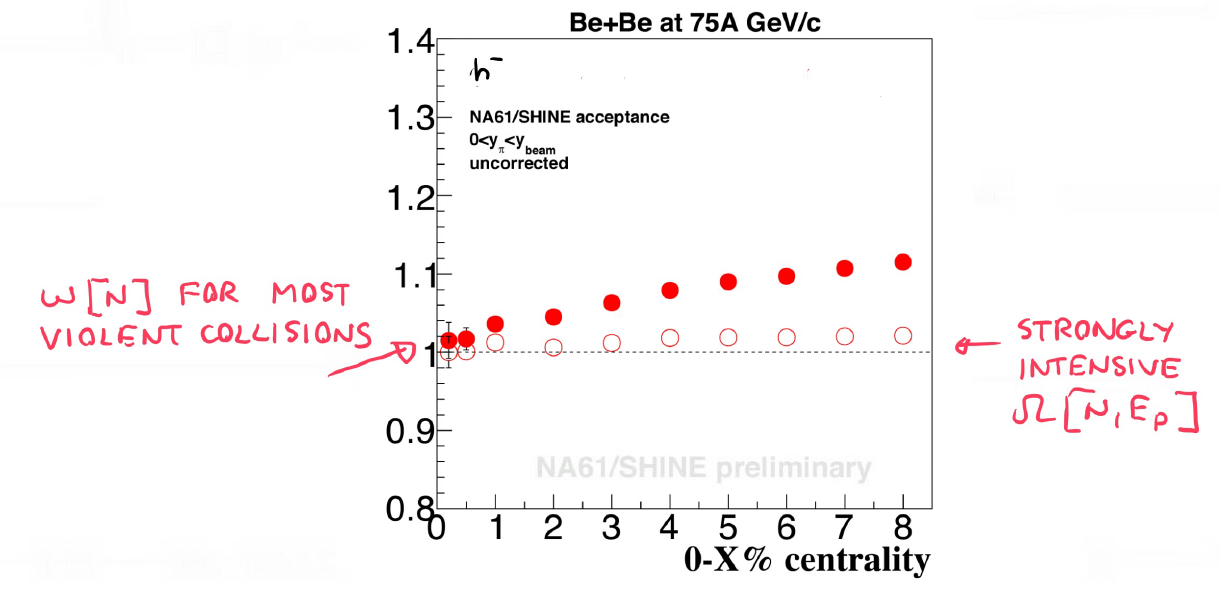}
	\caption[]
	{
		Comparison of the scaled variance of the negatively charged hadron multiplicity distribution with the corresponding
		strongly intensive scaled variance for Be+Be collisions at 75$A$~GeV/$c$ for different selections of violent
		(central) collisions. For detail see Ref.~\cite{cpod-seryakov}.
	}
	\label{fig:omega.Omega}
\end{figure}

The above defined procedures to minimize the effects of \textit{volume} fluctuations were tested using theoretical models and
experimental data. An exmaple of the experimental test performed for Be+Be collisions at 75$A$~GeV/$c$ is presented in Fig.~\ref{fig:omega.Omega}. The scaled variance $\omega$ increases with increasing percentile of violent collisions selected 
for the analysis using the measured values of $E_F$.
The strongly intensive scaled variance $\Omega$ is almost independent of the percentile of selected collisions. 
The two coincide for the most violent collisions.

\clearpage

\section{Exploiting conservation laws}
\label{sec:conservation}

\begin{figure}[h]
	\centering
	\includegraphics[width=0.90\textwidth]{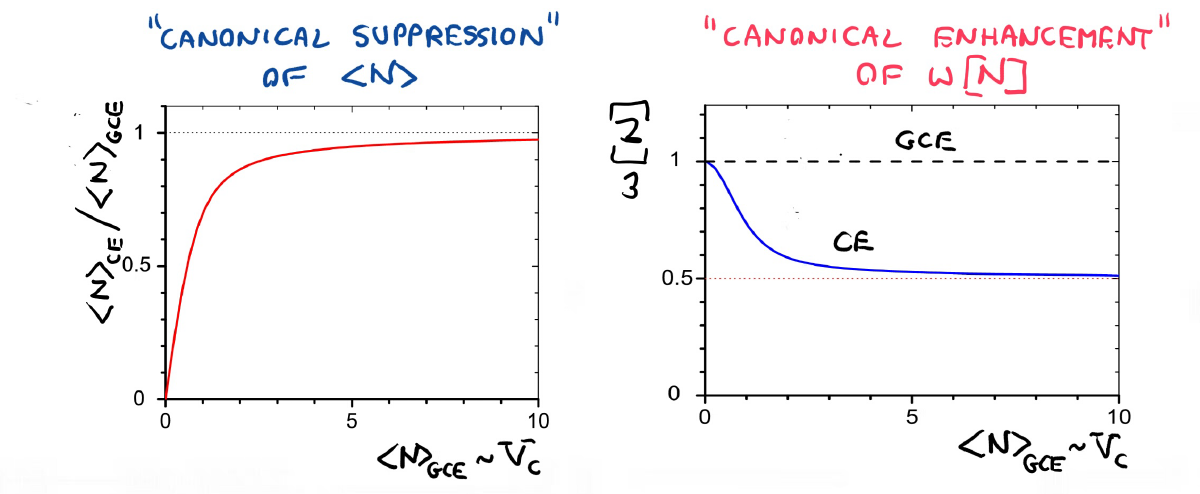}
	\caption[]
	{
		The ratio of  mean multiplicities  
		(\textit{left}) and scaled variance (\textit{right})
		calculated for the SM(IB-CE) and SM(IB-GCE) ensembles are 
		plotted as a function of the mean multiplicity for SM(IB-GCE), the latter 
		being proportional to the cluster volume, $V_C$.
		The calculations are done for a gas of positively and negatively
		charged particles with total charge equal to zero.
	}
	\label{fig:canonical}
\end{figure}

The most efficient description of bulk properties of hadron production in high energy heavy ion collisions
is given by statistical and hydrodynamical models~\cite{florkowski}.
Thus, here the impact of conservation laws on mean multiplicity and multiplicity fluctuations is
discussed with statistical models.
For simplicity consider only material  conservation laws (conserved charges: Q, B, S, ...).
Then the simplest model is a Statistical Model with Ideal Boltzmann Gas using the Canonical Ensemble, SM(IB-CE).
In this model correlations between particles are only due to material conservation laws.
Their influence on mean multiplicity and scaled variance is presented in Fig.~\ref{fig:canonical}, where
results for a gas of positively and negatively charged particles contained in a cluster of volume $V_C$ and
with total charge equal to zero are shown as
a function of mean multiplicity calculated with the  SM(IB-GCE). Note that the mean multiplicity in the SM(IB-GCE) is proportional
to the cluster volume provided the temperature is independent of $V_C$.
The well know ''canonical suppression'' of the mean multiplicity~\cite{danos} and
a somewhat less well known ''canonical enhancement'' of the scaled variance~\cite{begun} are seen. 
Note that in the large $V_C$ limit the results for the mean multiplicity obtained with the SM(IB-CE) and SM(IB-GCE) are
equal and for the scaled variance they remain different.  
  
\begin{figure}[h]
	\centering
	\includegraphics[width=0.45\textwidth]{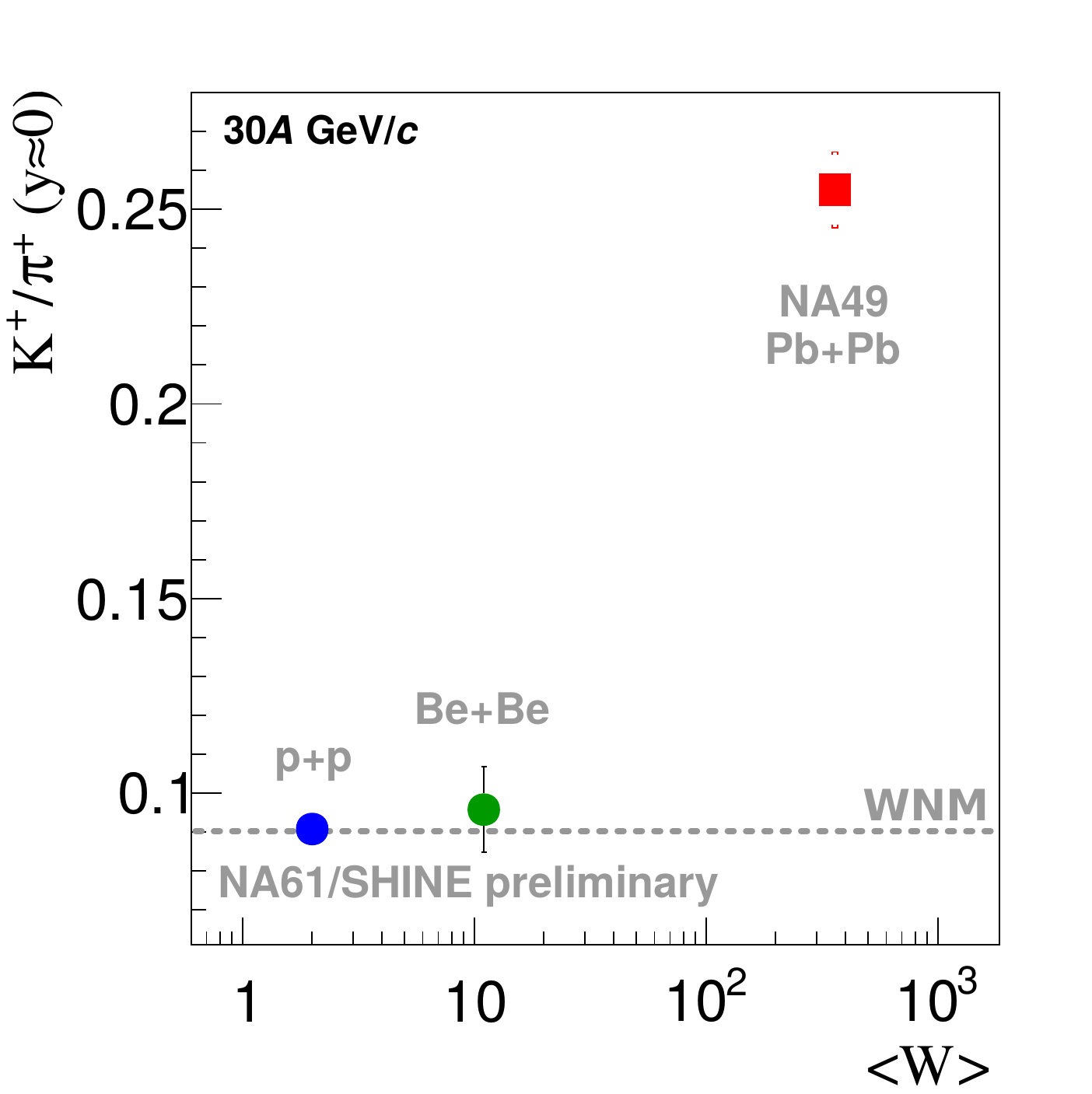}
	\includegraphics[width=0.45\textwidth]{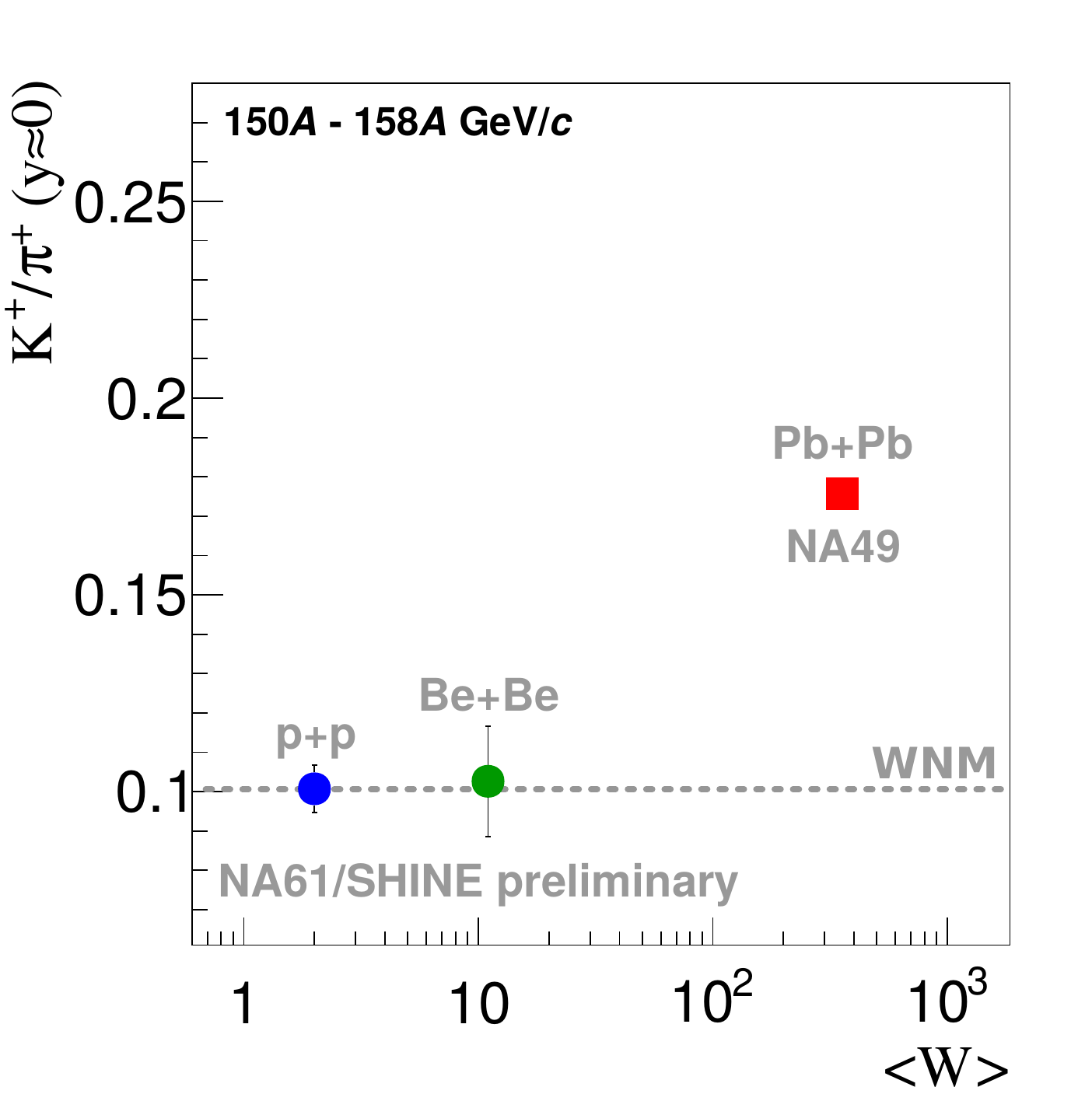}
	\centering
	\hspace*{0.9cm}\includegraphics[width=0.45\textwidth]{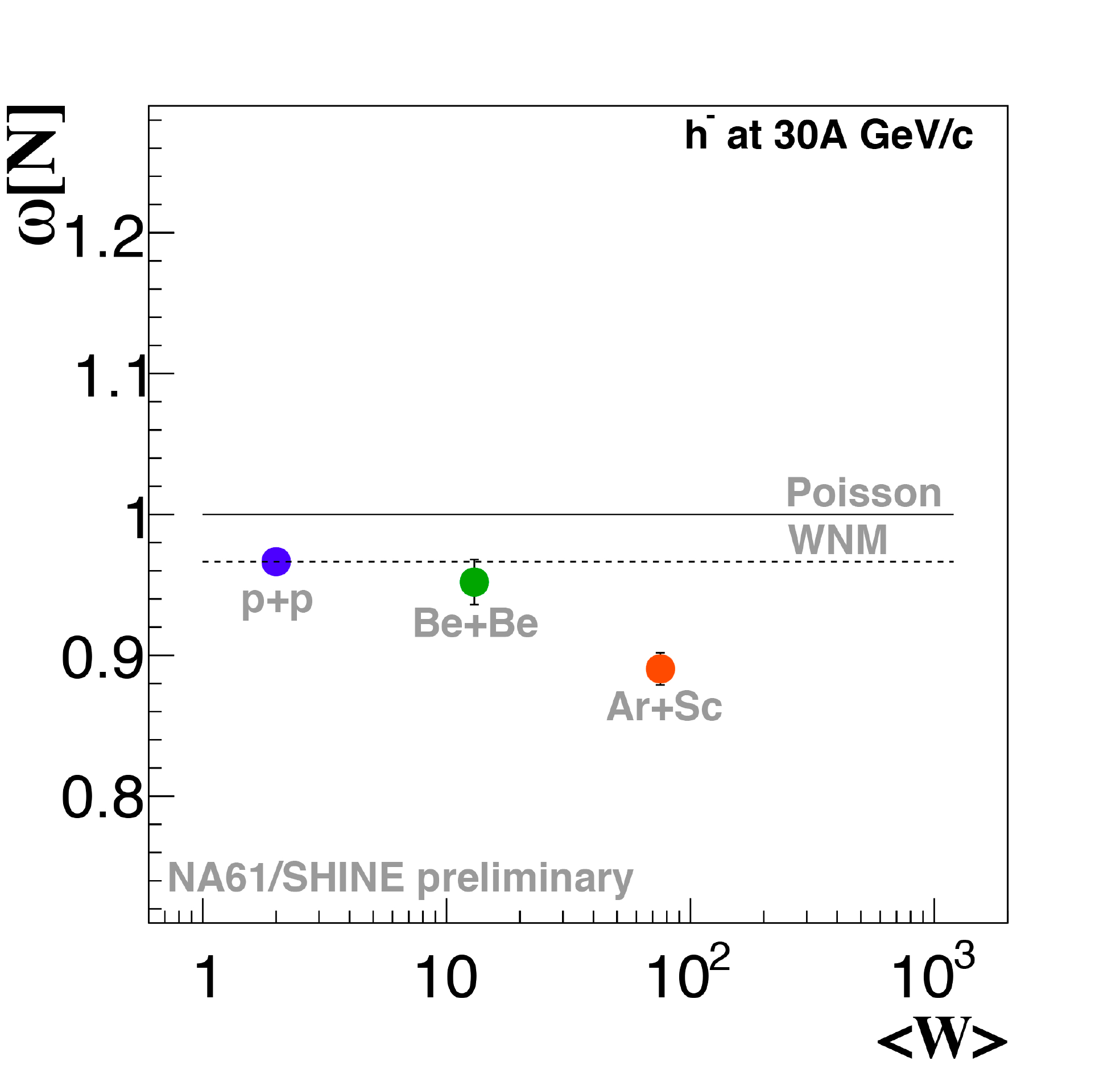}
	\includegraphics[width=0.45\textwidth]{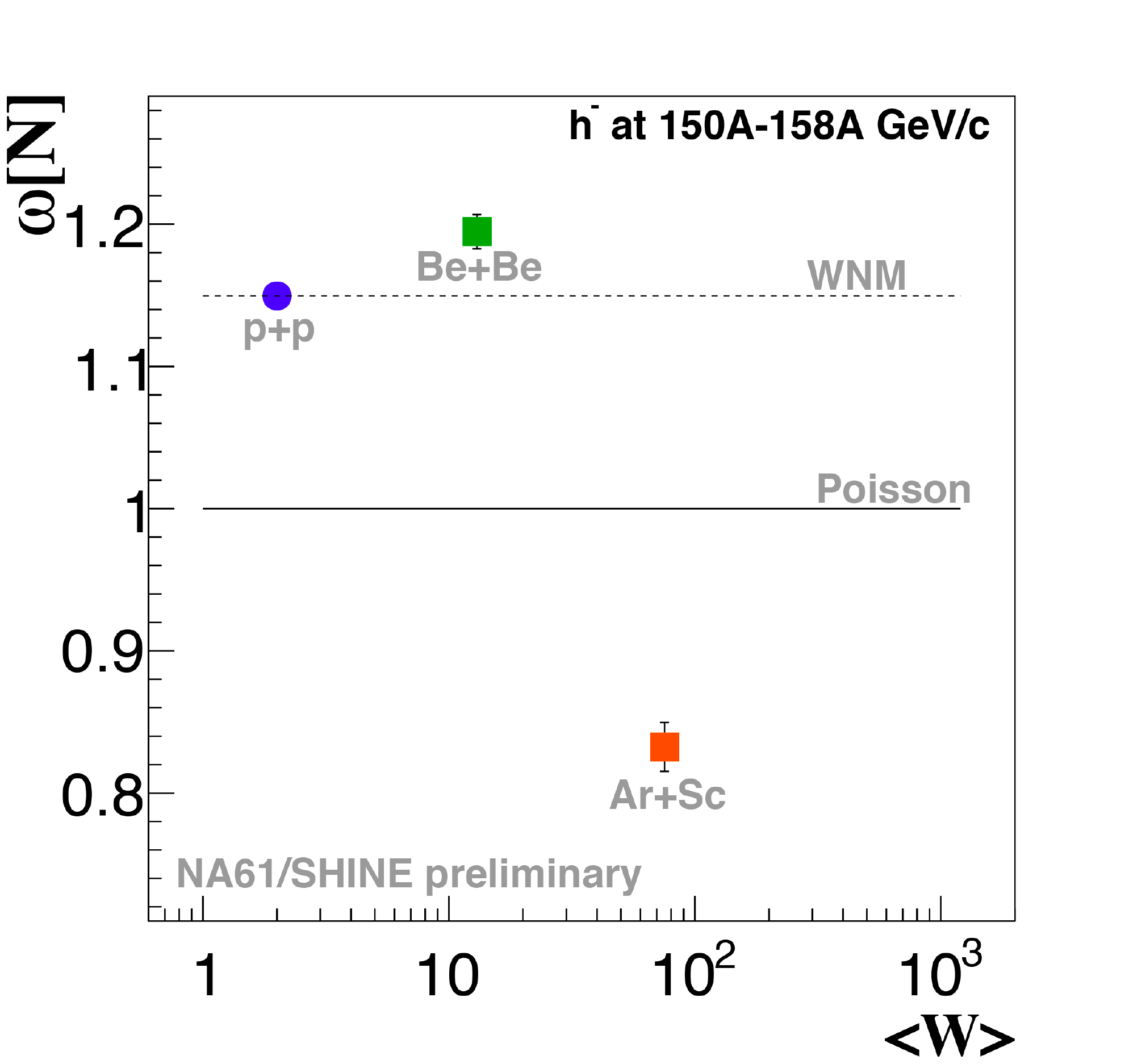}
	\caption[]{
		\textit{Top left} and \textit{top right}: 
		System size dependence of the $K^{+}/\pi^{+}$ ratio at mid-rapidity at 30$A$~GeV/$c$ and 150$A$~GeV/$c$. 
		\textit{Bottom left} and \textit{bottom right}: 
		System size dependence of the scaled variance of the multiplicity distribution of negatively 
		charged hadrons at 30$A$~GeV/$c$ and 150$A$~GeV/$c$.}
	\label{fig:OoF}
\end{figure}

In the following the qualitative expectations derived from the SM(IB-CE) will be confronted with experimental data.
Figure~\ref{fig:OoF} shows example plots on the system size dependence of the ratio
of $K^+$ and $\pi^+$ yields at mid-rapidity and of the scaled variance of multiplicity distributions. 
The Be+Be results are  close to p+p independently of collision energy. 
Moreover, the data show a jump between light (p+p, Be+Be) and intermediate, heavy (Ar+Sc, Pb+Pb) 
systems.
 
\begin{figure}[h]
	\centering
	\includegraphics[width=0.90\textwidth]{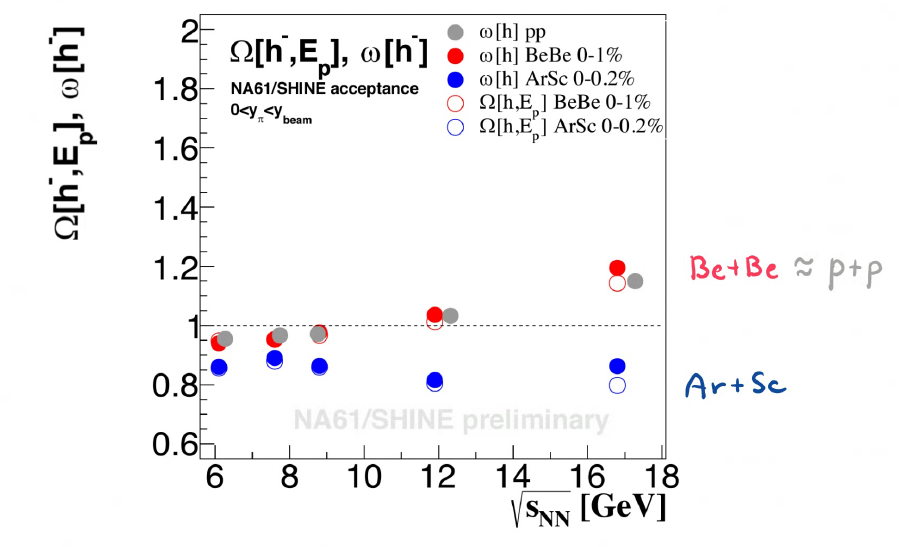}
	\caption[]
	{
		Collision energy dependence of scaled variance and strongly intensive scaled variance of the negatively charged hadron multiplicity distribution for inelastic p+p interactions and violent Be+Be and
		Ar+Sc collisions.
	}
	\label{fig:omegaVSenergy}
\end{figure}

Here one recalls the following:
\begin{enumerate}
	\item
	The $K^+/\pi^+$ ratio in p+p interactions is below the 
	predictions of statistical models. However, the ratio in central Pb+Pb collisions is  
	close to statistical model predictions for large volume systems. 
	For detail see e.g. Ref.~\cite{Becattini:2005xt} 
	\item
	In p+p interactions, and thus also in Be+Be collisions, multiplicity fluctuations are larger than predicted by statistical models.
	However, they are close to statistical model predictions for large volume systems in central Ar+Sc and Pb+Pb collisions,
	for detail see Ref.~\cite{Begun:2006uu}.
\end{enumerate}
Thus the observed rapid change of hadron production properties that start when moving from Be+Be 
to Ar+Sc collisions can be interpreted as the beginning of creation of large clusters of 
strongly interacting matter - the onset of fireball~\cite{sr2017}. 
We note that non-equilibrium clusters produced in p+p and Be+Be collisions seem to have similar properties
at all beam momenta studied here.
This is well seen in Fig.~\ref{fig:omegaVSenergy} where scaled variance and strongly intensive scaled variance are plotted
as a function of collision energy.

\begin{figure}[h]
\centering
\includegraphics[width=0.90\textwidth]{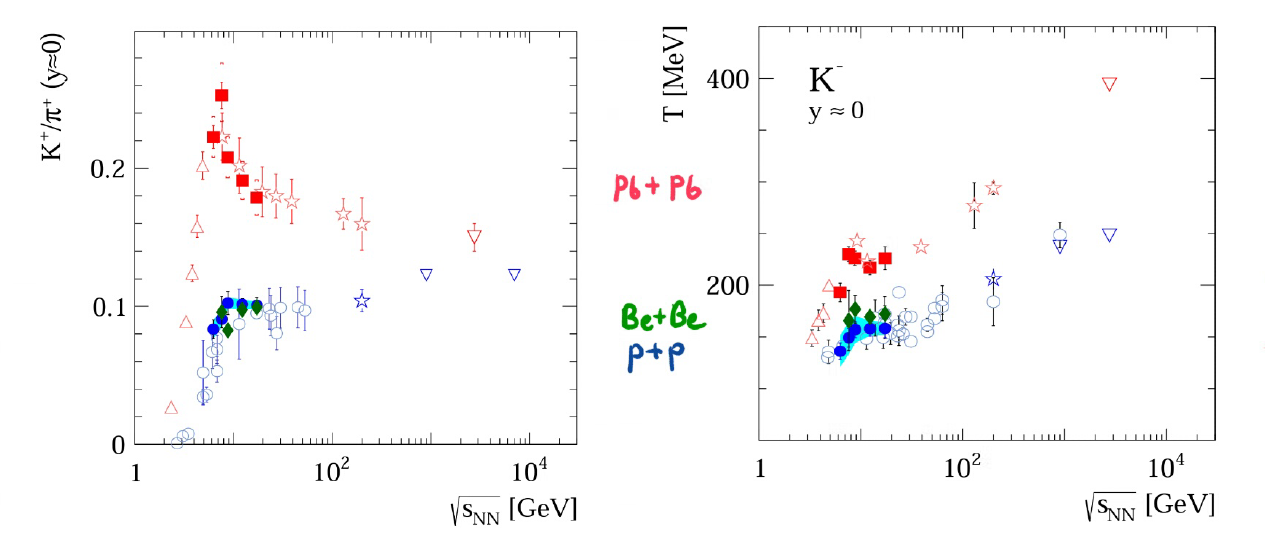}
\caption[]
{
\textit{Left:} Energy dependence of positively charged kaon yield divided by the corresponding charged pion yield at mid-rapidity.
\textit{Right:} Energy dependence of inverse slope parameter of transverse mass spectra of negatively charged kaons at mid-rapidity.
Results for inelastic p+p interactions and violent Be+Be and Pb+Pb (Au+Au) collisions are presented.	
}
\label{fig:HornStep}
\end{figure}

Finally we recall that hadron production properties in heavy ion collisions were found to change 
rapidly with increasing
collision energy in the low SPS energy domain, $\sqrt{s_{NN}} \approx 10$~GeV
(for a recent review see Ref.~\cite{Gazdzicki:2014sva}). The NA61/SHINE results shown in
Fig.~\ref{fig:HornStep} indicate that this is also the case in 
inelastic p+p interactions and probably also in Be+Be collisions. The phenomenon is labelled
as the \textit{onset of deconfinement} and interpreted as the beginning of creation of
quark-gluon plasma with increasing collision energy~\cite{Gazdzicki:2010iv}. 

\begin{figure}[h]
	\centering
	\includegraphics[width=0.90\textwidth]{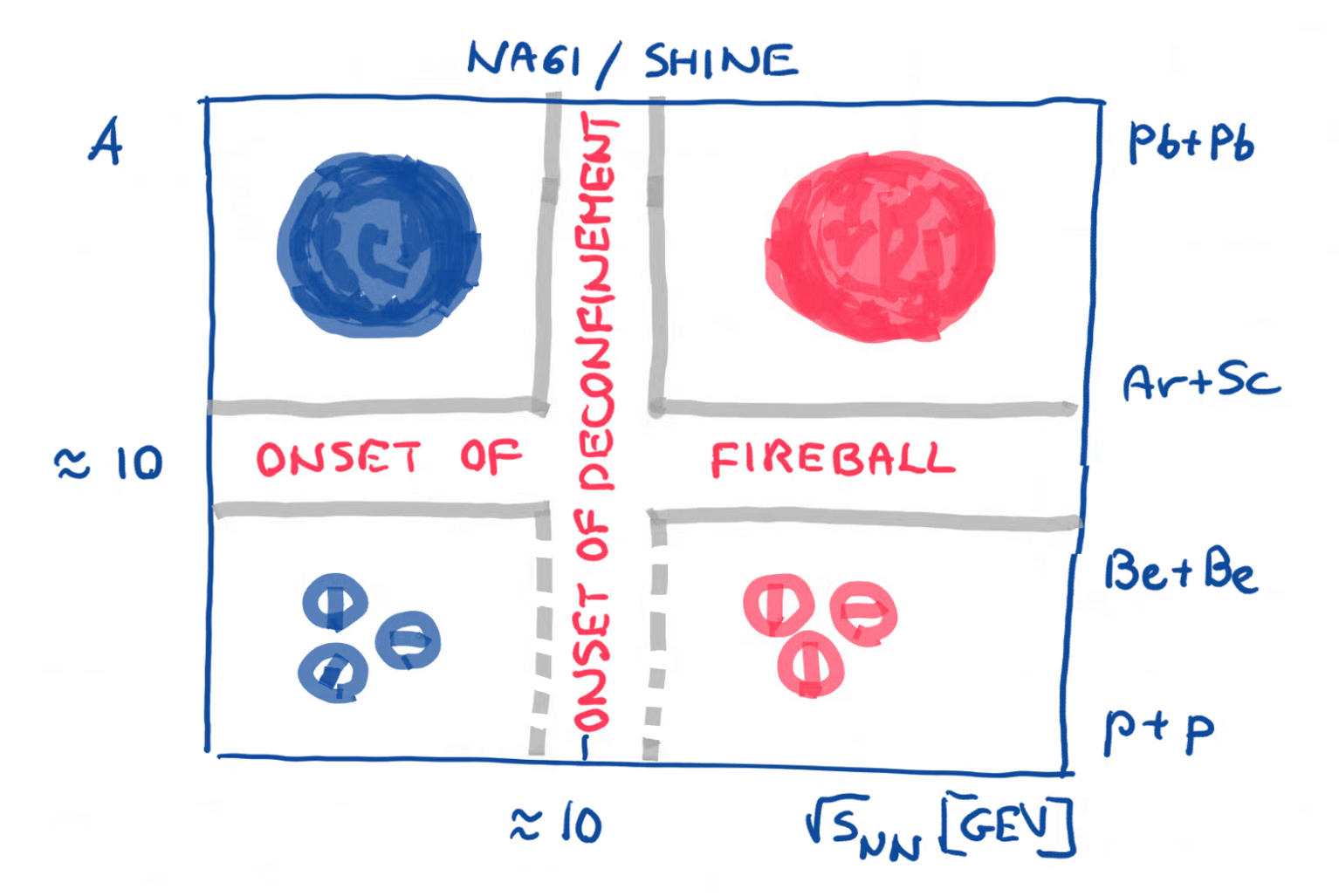}
	\caption[]
	{
	Two-dimensional scan conducted by NA61/SHINE by varying collision energy and nuclear
	mass number of colliding nuclei indicates four domains of hadron production properties
	separated by two thresholds: the onset of deconfinement and the onset of fireball.
	The onset of deconfinement is well established in central Pb+Pb(Au+Au) collisions,
	its presence in collisions of low mass nuclei, in particular, inelastic p+p interactions
	is questionable.
 	}
	\label{fig:cross}
\end{figure}

Consequently the two-dimensional scan conducted by NA61/SHINE by varying collision energy and nuclear
mass number of colliding nuclei indicates four domains of hadron production properties
separated by two thresholds: the onset of deconfinement and the onset of fireball.
The sketch presented in Fig.~\ref{fig:cross} illustrates this conclusion.

\clearpage

\section{Search for critical point}
\label{sec:critical}

A characteristic feature of a second order phase transition
(the critical point or line) is the
divergence of the correlation length.
The system becomes scale invariant. 
This leads to large fluctuations in particle multiplicity.
Moreover these fluctuations have specific 
characteristics~\mbox{\cite{Wosiek:1988,Bialas:1990xd}}.  
Also other properties of the system should be sensitive to the
vicinity of the critical point~\cite{Stephanov:1999zu}.
Thus when scanning the phase diagram a region of increased  
fluctuations may signal the critical point
or the critical line.

In the experimental search for the critical point one would like to minimize influence of both \textit{volume} fluctuations 
and conservation laws. Thus it is recommended to use quantities which are insensitive (in the WNM and the SM(IB-GCE)) to these.
It appears that strongly intensive quantities composed of suitably selected extensive quantities have this property.
For example these are~\cite{siq,Gazdzicki:2013ana}:
\begin{equation}
\Delta[P_{T},N] =
\frac {1}{\langle N \rangle \omega[p_{T}]}[\langle N \rangle \omega[P_{T}] - 
\langle P_{T} \rangle \omega[N]]
\label{eq:delta1}
\end{equation}
and
\begin{equation}
\Sigma[P_{T},N] =
\frac{1}{\langle N \rangle \omega[p_{T}]}[\langle N \rangle \omega[P_{T}] +
\langle P_{T} \rangle \omega[N] - 2(\langle P_{T}N \rangle - 
\langle P_{T} \rangle \langle N \rangle )]\ ,
\label{eq:sigma1}
\end{equation}
where $P_{T}$ is the sum of the absolute
values of transverse momenta $p_{T}$. 
The quantity $\omega[p_{T}]$ is the scaled variance of
the inclusive $p_{T}$ distribution (summation runs over all particles and all events)
\begin{equation}
\omega[p_{T}] =
\frac{\overline{p_{T}^2} - \overline{p_{T}}^2}{\overline{p_{T}}}.
\end{equation}
It is easy to show that
$\Delta[P_{T},N]$ and $\Sigma[P_{T},N]$ are independent of \textit{volume} fluctuations and material conservation laws for the SM(IB-CE).

\begin{figure}[h]
	\centering
	\includegraphics[width=0.90\textwidth]{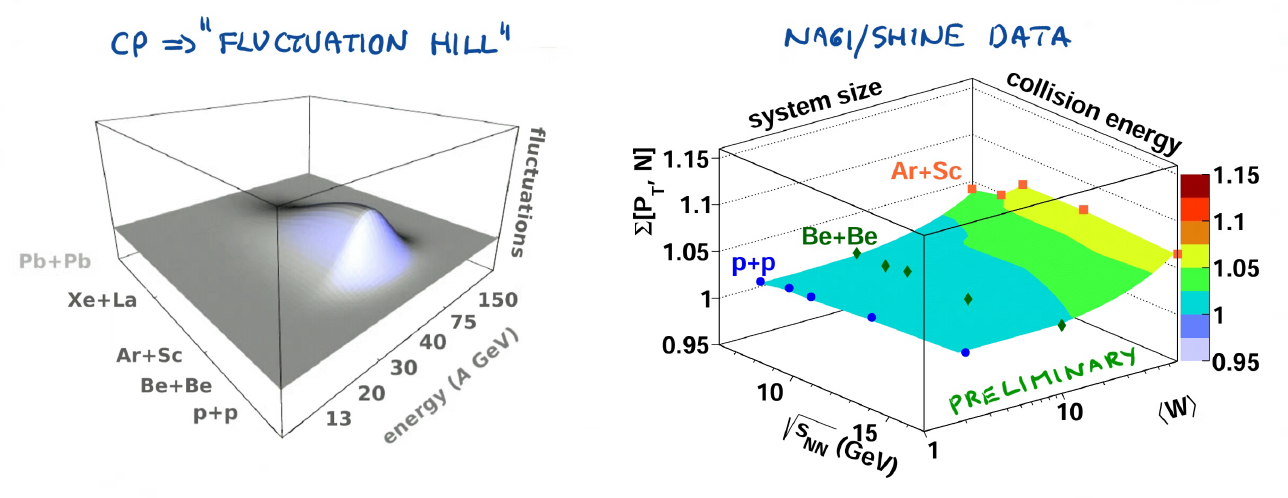}
	\caption[]
	{
    \textit{Left}: Sketch of the hill of fluctuations which may be observed in the (beam momentum) - (system size) scan of
    NA61/SHINE provided the freeze-out parameters are close to the critical point.
    \textit{Right}: $\Sigma[P_{T},N]$ measured by NA61/SHINE in inelastic p+p interactions and violent Be+Be and Ar+Sc collisions
    at the CERN SPS energies. Results refer to negatively charged hadrons at forward rapidity ($0 < y_{\pi} < y_{beam}$) 
    and $p_T < 1.5$~GeV/$c$.
	}
	\label{fig:hill}
\end{figure}

Example results of NA61/SHINE from the search for the critical point using $\Delta[P_{T},N]$ and $\Sigma[P_{T},N]$
are presented in Fig.~\ref{fig:hill}. No indication for the critical point is observed so far.

In the grand canonical ensemble the correlation length $\xi$  diverges
at the critical point (or second order phase transition line) and
the system becomes scale invariant~\cite{Wosiek:1988,Satz:1989vj}. 
This leads to large multiplicity fluctuations with special properties.
They can be conveniently exposed using scaled 
factorial moments $F_r(M)$~\cite{Bialas:1985jb} of rank (order) $r$: 

\begin{equation}
F_r(M)=\frac{ \langle \displaystyle{\frac{1}{M}\sum_{i=1}^{M}} 
	N_i(N_i-1)...(N_i-r+1) \rangle }
{\langle \displaystyle{\frac{1}{M}\sum_{i=1}^{M}} N_i \rangle^r } ~,
\label{eq:facmom}
\end{equation}
where
$M = \Delta / \delta$ is the number of the subdivision intervals of size
$\delta$ of the momentum phase space region $\Delta$.
$N_i$ refers to particle multiplicity in the interval $i$ and
$\langle ... \rangle$ indicates averaging over the analysed collisions.

In the SM(IB-GCE) one gets $F_r(M) = 1$
for all values of $r$ and M provided 
the mean particle multiplicity is proportional to $\delta$.
The latter condition is trivially obeyed for a subdivision in configuration
space where the particle density is uniform throughout the
gas volume. For the case of subdivision in momentum
space the subdivision should be performed using so-called 
cumulative kinematic variables~\cite{Bialas:1990dk} in which 
the particle density is uniform.

At the second order phase transition the matter properties 
strongly deviate from the ideal gas.  The system is a simple
fractal and $F_r(M)$ possess
a power law dependence on $M$:
\begin{equation}
F_r(M) = F_r(1) \cdot M^{-\phi_r} ~.
\label{eq:p1}
\end{equation}
Moreover the exponent (intermittency index) $\phi_r$ satisfies the relation:
\begin{equation}
\phi_r = ( r - 1 ) \cdot d_r ~,
\label{eq:p2}
\end{equation}
with the anomalous fractal dimension $d_r$ being independent 
of $r$~\cite{Bialas:1990xd}.

Note, that $F_r(M)$ is sensitive to to both \textit{volume} fluctuations and material conservation laws.
A formulation of a new method to study intermittency using strongly intensive quantities is needed.

\begin{figure}[h]
	\centering
	\includegraphics[width=0.90\textwidth]{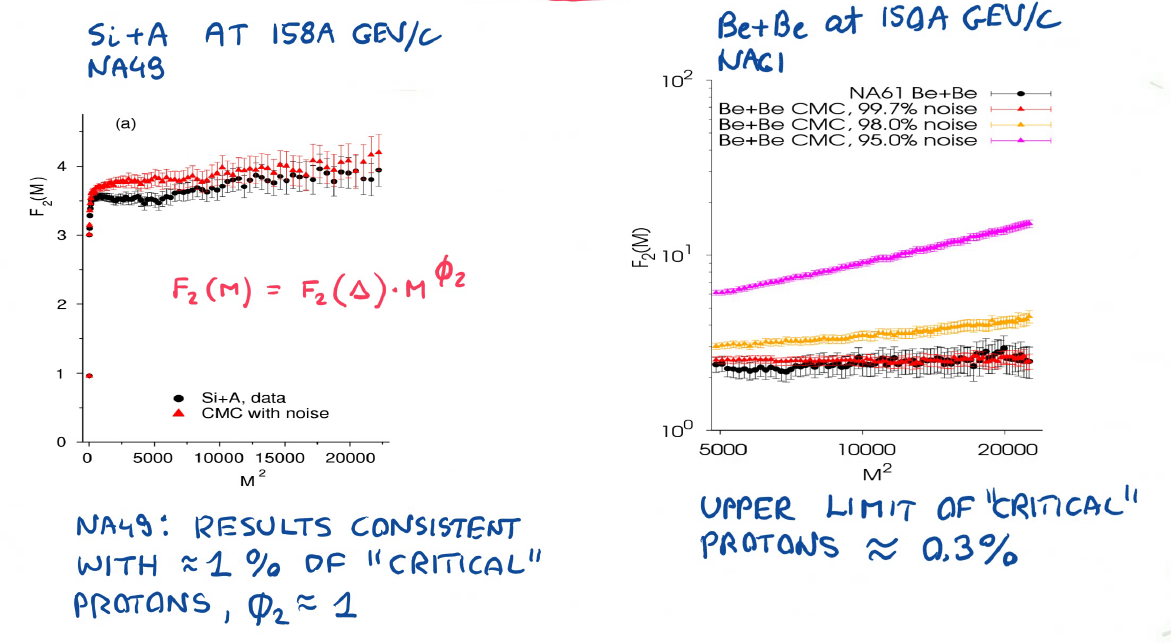}
	\caption[]
	{
    $F_2(M)$ of protons in violent Si+A collisions at 158$A$~GeV/$c$ (\textit{left}) and in violent Be+Be collisions 
    at 150$A$~GeV/$c$ (\textit{right}). The subdivision is performed in transverse momentum plane at mid-rapidity. 
    The results are compared to model predictions calculated for different fraction of protons obeying critical fluctuations. 
	}
	\label{fig:intermittency}
\end{figure}

NA61/SHINE results and NA49 results on $F_2(M)$ for protons are presented in Fig.~\ref{fig:intermittency}.
The NA49~\cite{Anticic:2012xb} results for violent Si+A collisions at 158$A$~GeV 
are consistent with about 1\% of protons obeying critical fluctuations 
with $\phi_2 \approx 1$.
The NA61/SHINE~\cite{sr2017} results for violent Si+A collisions at 150$A$~GeV/$c$ establish 
the upper limit for protons obeying critical fluctuations 
to be about 0.3\%. For details see Ref.~\cite{davis-cpod}.

\clearpage

%\section{Closing remarks}

\end{document}